\documentclass[%
preprint,
 amsmath,amssymb,
pra,
]{revtex4-1}

\usepackage{graphicx}
\usepackage{dcolumn}
\usepackage{bm}

\usepackage{CJK}

\usepackage{color}
\usepackage{amssymb}
\usepackage{amsfonts}

\usepackage[colorlinks,urlcolor=blue,linkcolor=blue,citecolor=blue,anchorcolor=blue]{hyperref}

\begin{document}


\preprint{APS/123-QED}

\title{Rabi oscillation of azimuthons in weakly nonlinear waveguides}

\author{Kaichao Jin$^{1,2}$}
\author{Yongdong Li$^1$}
\author{Feng Li$^1$}
\author{Milivoj R. Beli\'c$^3$}
\author{Yanpeng Zhang$^{1}$}
\author{Yiqi Zhang$^{1,2,}$}
\email[Corresponding author: ]{zhangyiqi@mail.xjtu.edu.cn}
\affiliation{%
 $^1$Key Laboratory for Physical Electronics and Devices of the Ministry of Education,
Xi'an Jiaotong University, Xi'an 710049, China \\
$^2$Guangdong Xi'an Jiaotong University Academy, Foshan 528300, China \\
$^3$Science Program, Texas A\&M University at Qatar, P. O. Box 23874 Doha, Qatar
}%

\date{\today}

\begin{abstract}
  \noindent Rabi oscillation, an inter-band oscillation, depicts the periodic flopping between two states that belong to different energy levels in the presence of an oscillatory driving field.
In photonics, Rabi oscillation can be mimicked by applying a weak longitudinal periodic modulation to the refractive index change of the system.
However, the Rabi oscillation of nonlinear states has yet to be discussed.
We report Rabi oscillations of azimuthons---spatially modulated vortex solitons---in weakly nonlinear waveguides with different symmetries, both numerically and theoretically.
The period of Rabi oscillation can be determined by applying the coupled mode theory,
which largely depends on the modulation strength.
Whether the Rabi oscillation between two states can be obtained or not is determined by the spatial symmetry of the azimuthons and the modulating potential.
In this paper we succeeded in obtaining the Rabi oscillation of azimuthons in the weakly nonlinear waveguides with different symmetries.
Our results not only enrich the Rabi oscillation phenomena, but also provide a new avenue in the study of pattern formation and spatial field manipulation in nonlinear optical systems.
\end{abstract}

\maketitle

\section{Introduction}

The Rabi oscillation originated in quantum mechanics \cite{rabi.pr.49.324.1936}, but by
now is much investigated in a variety of optical and photonic systems that include fibers \cite{hill.el.26.1270.1990,lee.ao.39.1394.2000},
multimode waveguides \cite{kartashov.prl.99.233903.2007,vysloukh.ol.40.4631.2015,zhang.oe.23.6731.2015},
coupled waveguides \cite{ornigotti.jpb.41.085402.2008},
waveguide arrays \cite{makris.oe.16.10309.2008,shandarova.prl.102.123905.2009,vysloukh.ol.39.5933.2014}, and
two-dimensional modal structures \cite{wong.science.337.446.2012,kartashov.ol.38.3414.2013}.
Recently, Rabi oscillations of topological edge states \cite{zhang.lpr.12.1700348.2018,zhong.ol.44.3342.2019} and modes
in fractional Schr\"odinger equation \cite{zhang.oe.25.32401.2017} were also reported.
Rabi oscillations are inter-band oscillations that require an ac field to be applied as an external periodic potential.
In optics, the longitudinal periodic modulation of the refractive index change plays the role of an ac field in temporal quantum systems,
and Rabi oscillations are indicated by the resonant mode conversion.
As far as we know, the investigation of optical Rabi oscillations thus far has been limited to the linear regime only,
and the Rabi oscillation in nonlinear systems is still an open problem that needs to be explored. It is addressed in this paper.

Hence, the aim of this work is to investigate Rabi oscillations of azimuthons in weakly nonlinear waveguides that is accomplished by applying a weak longitudinally modulated periodic potential.
Azimuthons are a special type of spatial solitons; they are azimuthally modulated vortex beams
that exhibit steady angular rotation upon propagation \cite{desyatnikov.prl.95.203904.2005}.
Generally, azimuthons, especially the ones with higher-order angular momentum structures, are unstable in media with local Kerr or saturable nonlinearities.
To overcome the instability drawback, a nonlocal nonlinearity is introduced, and recently published reports demonstrate that
the stable propagation of azimuthons can indeed be obtained \cite{lopez-aguayo.oe.14.7903.2006,buccoliero.prl.98.053901.2007,buccoliero.ol.33.198.2008}.
In addition, it was also reported that the spin-orbit-coupled Bose-Einstein condensates can support stable azimuthons as well \cite{kartashov.prl.122.123201.2019}.
However, the treatment of nonlocal nonlinearity and spin-orbit-coupled Bose-Einstein condensates is challenging in both theoretical modeling and experimental demonstration.
Nevertheless, it has been confirmed that the weakly nonlinear waveguides \cite{stefan.pre.70.016614.2004} represent an ideal platform for the investigation of stable azimuthons \cite{zhang.oe.18.27846.2010,thomax.lpr.6.709.2012},
even with higher-order modal structures.

Following this path of inquiry, we first investigate Rabi oscillations of azimuthons in a circular waveguide and then in a square waveguide.
Since in this nonlinear three-dimensional wave propagation problem no analytical solutions are known, necessarily the mode of inquiry will be predominantly numerical with some theoretical background.
In the circular waveguide, the azimuthons will exhibit Rabi oscillation while rotating during propagation.
In the square waveguide, the behavior of the azimuthons is different in two aspects \cite{zhang.oe.25.32401.2017}:
(i) azimuthons will rotate only if the corresponding Hamiltonian (energy) is bigger than a certain threshold value;
(ii) azimuthons will be deformed during propagation.
Hence, in this work we choose azimuthons with large enough energies to avoid wobbling motions in the square waveguide during propagation.

\section{Results}

\subsection{Theoretical analysis}
The propagation of a light beam in a photonic waveguide can be described by the Schr\"odinger-like paraxial wave equation
\begin{equation}\label{eq1}
  i \frac{\partial}{\partial Z} \Psi +\frac{1}{2 k_{0}}\left(\frac{\partial^{2}}{\partial X^{2}}+\frac{\partial^{2}}{\partial Y^{2}} \right) \Psi + k_{0} \frac{n_{2}}{n_{b}}|\Psi|^{2} \Psi
  +k_{0} \frac{n(X, Y)-n_{b}}{n_{b}} \left[1+\mu\cos(\delta Z)\right] \Psi=0,
\end{equation}
where $\Psi(X, Y, Z)$ is the complex amplitude of the light beam, the quantities
$(X, Y)$ and $Z$ are the transverse and longitudinal coordinates,
and $k_0=2 \pi n_b / \lambda_0$ with $\lambda_0$ being the wavelength.
The other quantities in Eq. (\ref{eq1}) are:
$\mu\ll1$ is the longitudinal modulation strength,
$\delta$ is the longitudinal modulation frequency,
$n_2$ is the nonlinear Kerr coefficient,
$n(X,Y)$ is the linear refractive index distribution, and
$n_b$ is the ambient index.
In Eq. (\ref{eq1}), the refractive index change includes two parts,
which are $|n-n_b|$ (linear part) and $n_2|\Psi|^2$ (nonlinear part).
We would like to note that weakly nonlinear waveguides demand not only
both the linear and nonlinear refractive index changes to be small in comparison with $n_b$,
but also the nonlinear part to be much smaller than the linear part.
According to the relations  $x=X/r_0$, $y=Y/r_0$, $z=Z/(k_0 r_0^2)$, $d=k_0 r_0^2 \delta$, and $\sigma=\mathrm{sgn}(n_2)$,
with $r_0$ being determined by the real beam width, Eq. (\ref{eq1}) can be rewritten into its dimensionless version
\begin{equation}\label{eq2}
i\frac{\partial}{\partial z} \psi + \frac{1}{2}\left( \frac{\partial ^2}{\partial x^2}
 + \frac{\partial ^2}{\partial y^2} \right) \psi
 + \sigma |\psi|^2 \psi + V[1+\mu\cos(d z)] \psi=0,
\end{equation}
with $\psi = k_0 r_0 \sqrt{|n_2|/n_b} \Psi$ and $V(x,y)=k_0^2 r_0^2 [n(x,y)-n_b]/n_b$.
Here, we will consider propagation in a deep circular potential
$
V(x,y) = V_0 \exp[-(x^2+y^2)^5/w^{10}]
$
with $w$ characterizing the potential width
and $V_0$ the potential depth.
To guarantee a weak nonlinearity, the potential should be deep enough.
Thus, the potential is deep but the potential modulation is shallow.
The parameter $\sigma =1$ $(\sigma =-1)$ corresponds to the focusing (defocusing) nonlinearity.
In this work, we consider the focusing nonlinearity, i.e., we take $\sigma =1$.

There are large amounts of materials to be used to produce waveguides, and
silica is one of the popular materials among them with
typical parameters $n_b=1.4$, $|n-n_b| \leq 9 \times 10^{-3}$, and $n_2 = 3\times 10^{-16} ~\mathrm{cm}^2/\mathrm{W}$
for light beams with wavelength ranging from visible to near-infrared.
Without loss of generality, we choose $\lambda_0=800\,\rm nm$ in this work.
Therefore if we choose $V_0=500$, the value of $r_0 \approx 25.0 ~\mu \mathrm{ m}$ can be obtained according to the relation
adopted in Eq. (\ref{eq2}).
Indeed, such a value is reasonable for a multi-mode fiber \cite{agrawal_book_2009}.
According to the wavelength and $r_0$, one knows that the diffraction length is $\sim 7\, \rm mm$.
Considering the group velocity dispersion coefficient is $\sim 35\,\rm fs^2/mm$ at the wavelength $\lambda_0=800\,\rm nm$,
the dispersion length is of the order of kilometers for a picosecond light beam,
which is much longer than the propagation distance taken in this work.
As a result, it is safely to neglect the temporal effect.

\begin{figure*}
  \includegraphics[width=\textwidth]{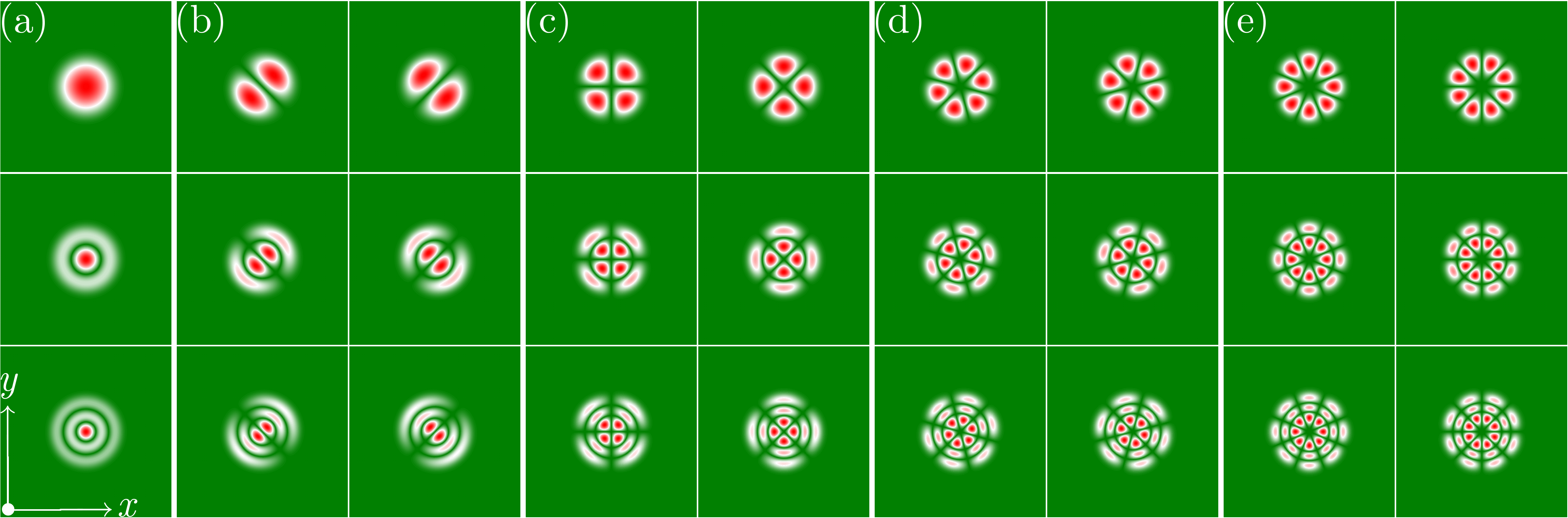}
  \caption{(a) Basic modes. (b) Degenerate dipole modes.
  (c) Degenerate quadrupole modes. (d) Degenerate hexapole modes. (e) Degenerate octopole modes.
  First row: first-order modes. Second row: second-order modes. Third row: third-order modes.
  The panels are shown in the window $-2 \le x \le 2$ and $-2 \le y \le 2$.
  Other parameters: $V_0=500$ and $w=1$.}
  \label{fig1}
\end{figure*}

To start with, we consider the modes supported by the deep potential alone, therefore
the nonlinear term and the longitudinal modulation in Eq. (\ref{eq2}) are initially neglected.
The corresponding solution of the reduced linear Eq. (\ref{eq2}) can be written as
$\psi(x,y,z)=u(x,y)\exp(i\beta z)$, with $u(x,y)$ being the stationary profile of the mode and $\beta$ the propagation constant.
Plugging this solution into the reduced Eq. (\ref{eq2}), one obtains
\begin{equation}\label{eq4}
\beta u = \frac{1}{2}\left( \frac{\partial ^2}{\partial x^2}
 + \frac{\partial ^2}{\partial y^2} \right) u + V u,
\end{equation}
which is the linear steady-state eigenvalue problem of Eq. (\ref{eq2}) with $\sigma$ and $\mu$ equal to zero.
Equation (\ref{eq4}) can be solved by utilizing the plane-wave expansion method,
and the eigenstates supported by the deep potential $V(x,y)$ can be easily obtained.
In Fig. \ref{fig1}, the first-order as well as higher-order basic modes, degenerate dipole modes,
degenerate quadrupole modes, degenerate hexapole modes, and
degenerate octopole modes that can exist in the potential are displayed. Here, the ``degenerate'' means  that all modes feature the same propagation constants,
in the usual optical meaning. These linear modes will be used as the input modes of the more general nonlinear and modulated modes of the complete Eq. (\ref{eq4}).

Therefore, to seek approximate azimuthons in weakly nonlinear waveguides, one takes the degenerate
modes and makes a superposition of them, as the initial wave
\begin{equation}\label{eq5}
\psi(x,y) = A [u_1(x,y)+iBu_2(x,y)] \exp(i\beta z),
\end{equation}
in Eq. (\ref{eq5}), where $A$ is an amplitude factor, $1-B$ the azimuthal modulation depth,
and $u_{1,2}(x,y)$ are the degenerate linear modes (see examples in Fig. \ref{fig1}).
Thus, we set the transverse profile of the azimuthon at the initial place as
\begin{equation}\label{eq6}
U(x,y,z=0)=A [u_1(x,y)+iBu_2(x,y)]
\end{equation}
and numerically propagate it, to obtain an output mode at arbitrary $z$.
We would like to note that
the inputs of the form (\ref{eq6}) do not rotate in linear medium,
since modes are degenerate. Rotation appears only when nonlinearity is added into model.

\begin{figure}\centering
  \includegraphics[width=0.5\columnwidth]{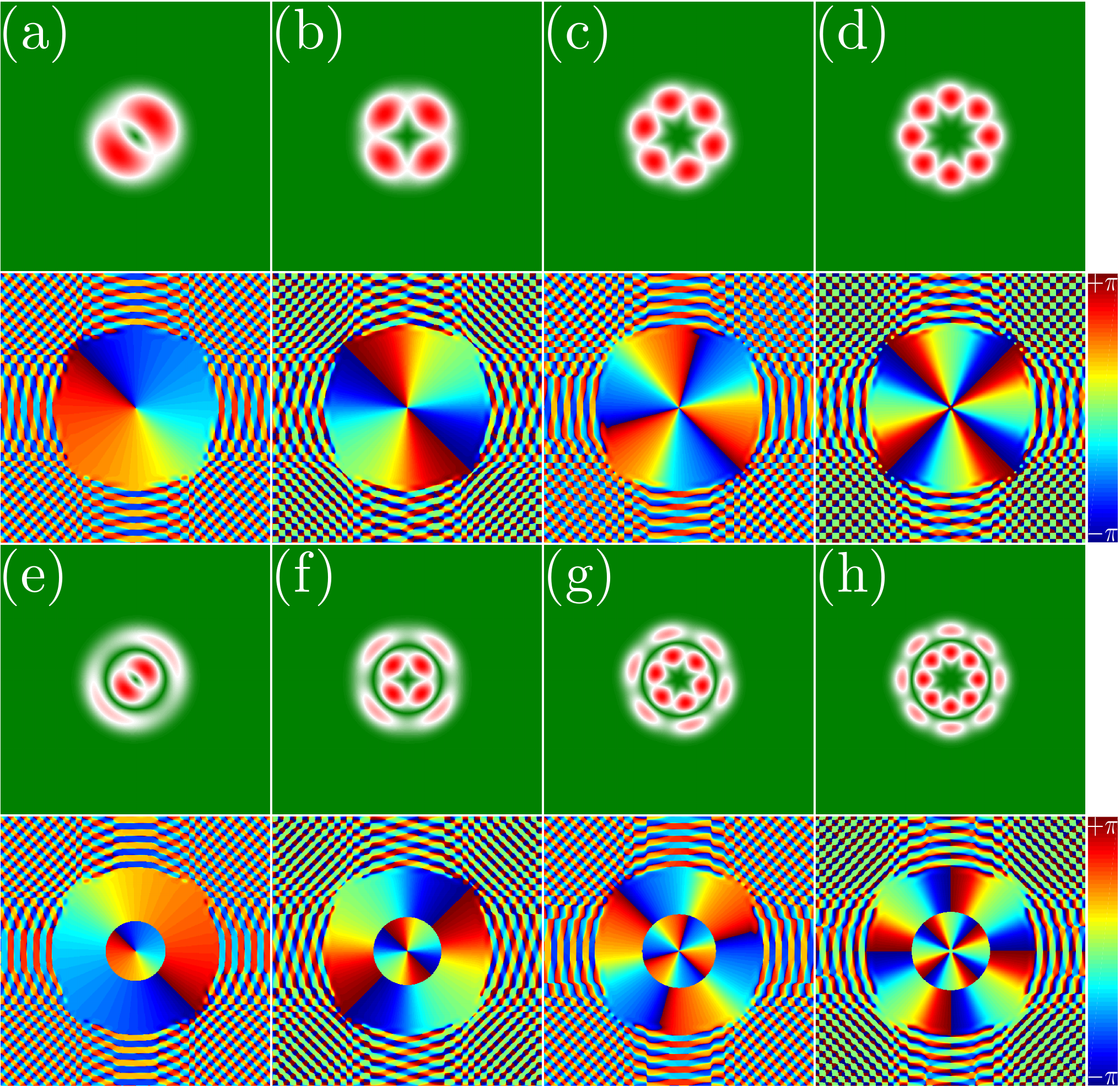}
  \caption{Amplitude and phase of the first-order (a-d) and second-order (e-h) azimuthons constructed from the degenerate dipoles (a,e),
  quadrupoles (b,f), hexapoles (c,g), and octopoles (d,h).
  The panels are shown in the window $-2 \le x \le 2$ and $-2 \le y \le 2$.
  Other parameters: $A=0.4$ and $B=0.5$.}
  \label{fig2}
\end{figure}

In Fig. \ref{fig2}, we display such approximate azimuthons with $A=0.4$ and $B=0.5$.
One finds that the phase of azimuthons is nontrivial, displaying
angular momentum and topological charge.
For dipole azimuthons, the topological charge is $\pm1$, while for
quadrupole, hexapole, and octopole azimuthons, the values are $\pm2$, $\pm3$ and $\pm4$, respectively.
As expected, these azimuthons will rotate with a constant frequency $\omega$ during propagation
when the nonlinear term in Eq. (\ref{eq2}) is included.
Therefore, the wave $U(x,y,z)$ can be rewritten as $U(r,\theta-\omega z)$ in polar coordinates,
with $r=\sqrt{x^2+y^2}$ and $\theta$ being the azimuthal angle in the transverse plane $(x,y)$. This fact allows for a bit of theoretical analysis.

After plugging Eq.~(\ref{eq5}) into Eq.~(\ref{eq2}) with $\mu=0$, multiplying by $U^*$ and $\partial_\theta U^*$ respectively,
and integrating over the transverse coordinates, one ends up with a linear system of equations
\begin{equation}\label{eq7}
\begin{split}
-\beta P + \omega L_z + I + N = &\,0, \\
-\beta L_z + \omega P' + I' + N' = &\,0,
\end{split}
\end{equation}
where
$
P = \iint |U|^2 dxdy,
$
$
L_{z}=-i\iint (- y \partial_x U + x \partial_y U) U^* d x d y,
$
$
P' = \iint |-y\partial_x U + x\partial_y U|^{2} dx dy,
$
$
I = \iint U^* \Delta_{\perp} U dx dy,
$
$
N = \iint [\sigma|U|^2+V ] |U|^2 dx dy,
$
$
I' = i\iint (-y \partial_x U^* + \partial_y U^*) \Delta_{\perp} U dx dy,
$
$
N' = i \iint (\sigma|U|^2 + V) (-y \partial_x U^* + x \partial_y U^* ) U d x d y.
$
Obviously, the quantities $P$ and $L_z$ stand for the power and angular momentum of the beam, and $P'$ is the norm of the state $\partial_\theta U$.
The integrals $I$ and $I'$ are related to the diffraction mechanism of the system,
while $N$ and $N'$ account for the waveguide and nonlinearity.
The angular frequency of the azimuthon during propagation can be obtained by directly solving Eq. (\ref{eq7}), that is
\begin{equation}\label{eq8}
\omega=\frac{P(I'+N')-L_z(I+N)}{L_z^2-PP'}.
\end{equation}

After these preliminaries, we are ready to address the Rabi oscillation of azimuthons.
To this end we adopt the superposition of two azimuthons $U_{m,n}(x,y)\exp(i\beta_{m,n}z)$ as an input
\begin{equation}\label{eq9}
    \psi=c_m(z)U_m(x,y)\exp(i\beta_m z)+c_n(z)U_n(x,y)\exp(i\beta_n z),
\end{equation}
where $c_{m,n}(z)$ are the slowly varying complex amplitudes of the azimuthons,
and the modulation frequency is $d=\beta_m-\beta_n$.
Plugging Eq. (\ref{eq9}) into Eq. (\ref{eq2}), and without considering the nonlinear term (but still on the level of analysis of Eq. (\ref{eq4})), one obtains
\begin{equation}\label{eq10}
\begin{split}
& i\frac{\partial c_m}{\partial z} U_m \exp(i d z) + \frac{1}{2}\mu c_m V U_m [1+\exp(2idz)] + \\
& i\frac{\partial c_n}{\partial z} U_n + \frac{1}{2}\mu c_n V U_n [\exp(idz)+\exp(-idz)] = 0.
\end{split}
\end{equation}
Since the azimuthons are constructed based on Eq. (\ref{eq6}),
they satisfy the relation $\langle U_m, U_n\rangle \neq 0$ if $m=n$ and  $\langle U_m, U_n\rangle = 0$ if $m\neq n$, thus forming a complete set of eigenstates.
Here, we borrowed the bra-ket notation from quantum mechanics.
Note that the orthogonality of azimuthon shapes is only valid in the weakly nonlinear regime.
As a result, one obtains two coupled equations based on Eq. (\ref{eq10})
\begin{equation}\label{eq11}
\begin{split}
i\frac{\partial c_m}{\partial z} + \frac{1}{2}\mu \frac{\langle U_mV U_n \rangle}{\langle U_m U_m \rangle} c_n & = 0,\\
i\frac{\partial c_n}{\partial z} + \frac{1}{2}\mu \frac{\langle U_nV U_m \rangle}{\langle U_n U_n \rangle} c_m & = 0,
\end{split}
\end{equation}
where $\langle U_{m} V U_{n}\rangle=\iint r U_{m}^{*} V U_n d r d\theta$, with the asterisk representing the conjugate operation.
Based on Eq. (\ref{eq11}), the period of Rabi oscillaiton can be obtained, as
\begin{equation}\label{eq12}
  z_{R}=\frac{\pi}{\left|\Omega_{R}\right|}
\end{equation}
with
\begin{equation}\label{eq13}
  \Omega_{R}=\frac{\mu }{2} \frac{\langle U_{m} V U_{n}\rangle}{\sqrt{\langle U_{m} U_{m}\rangle\langle U_{n} U_{n}\rangle}}.
\end{equation}
We note that the azimuthon conversion happens at half of the period, i.e., at $z_R/2$. Note also that the Rabi spatial frequency directly depends on the modulation strength $\mu$.

\subsection{Circular waveguide}

\begin{figure}\centering
  \includegraphics[width=0.5\columnwidth]{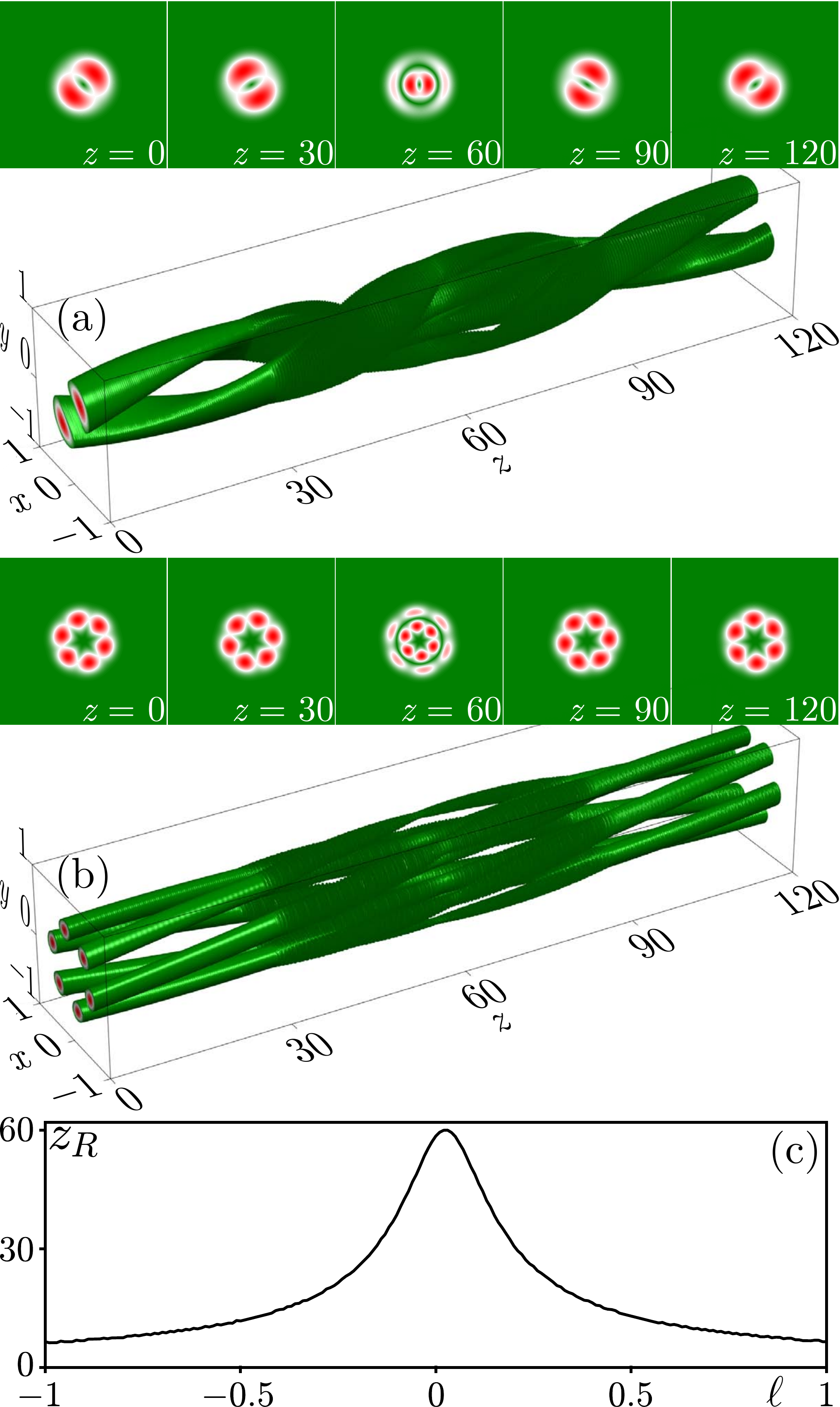}
  \caption{Rabi transition of a dipole (a) and a hexapole (b). In each case, the propagation is shown by the iso-surface plot, above which amplitude distributions at selected distances are shown.
  In both cases, the weak longitudinally periodic modulation exists in the region $30\le z\le 90$ with
  $d\approx25.2$ and $\mu\approx0.031$ in (a), and $d\approx36.36$ and $\mu\approx0.014$ in (b).
  (c) Rabi oscillation period $z_R$ versus frequency detuning $\ell$.}
  \label{fig3}
\end{figure}

We investigate the propagation of azimuthons in the circular weakly nonlinear waveguide by also including the longitudinal modulation,
and the results are displayed in Fig. \ref{fig3}.
Without loss of generality, we choose the dipole and hexapole azimuthons, which are shown in Fig. \ref{fig3}(a) and \ref{fig3}(b), respectively.
By taking the dipole azimuthon as an example [Fig. \ref{fig3}(a)],
we want to see whether the Rabi oscillation between the dipole azimuthon [Fig. \ref{fig2}(a)] and its corresponding second-order dipole azimuthon [Fig. \ref{fig2}(e)] can be established.
So, to induce resonance, we set the modulation frequency to be the difference between the eigenvalues of the two modes, which is $d \approx 25.2$.
As shown by Eq. (\ref{eq13}), the period of the Rabi oscillation is expected to depend on the modulation strength $\mu$,
and here we set it to be $\mu\approx 0.031$, to make the period $z_R\sim 60$.
As a consequence, one expects to see the second-order dipole azimuthon at a distance $\sim 30$ after turning on the longitudinal modulation.

\begin{figure}\centering
  \includegraphics[width=0.5\columnwidth]{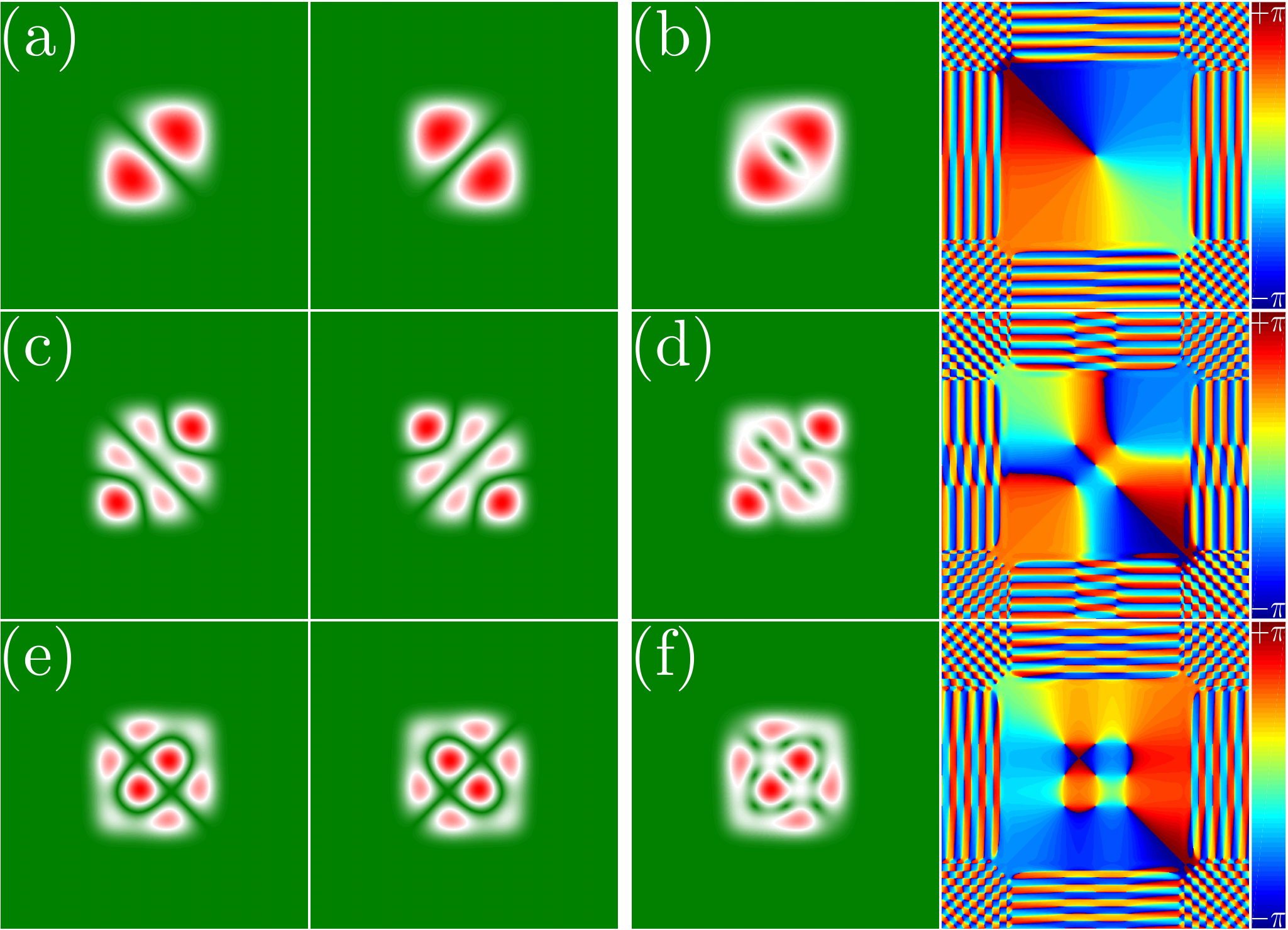}
  \caption{(a) Rabi transition of a deformed dipole. (b) Amplitude and phase of the azimuthon based on the dipole in (a).
  (c) Transition of a deformed hexapole. (d) Amplitude and phase of the azimuthon based on the hexapole in (c).
  (e) Transition of a deformed higher-order dipole. (f) Amplitude and phase of the azimuthon based on the higher-order dipole in (e).
  The panels are shown in the window $-2 \le x \le 2$ and $-2 \le y \le 2$.
  Other parameters: $A=0.4$ and $B=0.5$.}
  \label{fig4}
\end{figure}

In Fig. \ref{fig3}(a), the propagation of the dipole azimuthon is exhibited as a 3D iso-surface plot,
in which the longitudinal modulation exists only in the interval $30\le z\le90$.
When the propagation distance is smaller than $z\le30$, one in fact observes the stable rotational propagation of the dipole azimuthon.
The selected amplitude distributions at $z=0$ and $z=30$ are shown above the 3D iso-surface plots.
In the interval $30\le z\le90$, which is about one period of the Rabi oscillation,
the oscillation between the dipole azimuthon and the second-order azimuthon is displayed,
in which the dipole azimuthon completely switches to the second-order azimuthon at $z=60$.
Indeed, the corresponding amplitude distribution is same as that in Fig. \ref{fig2}(e) except for a rotation,
and the reason is quite obvious---azimuthons rotate steadily during propagation.
When the propagation distance reaches $z=90$, the dipole azimuthon is recovered and the longitudinal modulation is also lifted at the same time.
Therefore, one observes a stable rotating dipole azimuthon in the interval $90\le z\le120$,
and the amplitude distributions at $z=90$ and $z=120$ which are dipole azimuthons explicitly,
are shown above the iso-surface plot.
The analogous propagation dynamics of the hexapole azimuthon is shown in Fig. \ref{fig3}(b),
the setup of which is same as that of Fig. \ref{fig3}(a); it
also clearly displayss the Rabi oscillation of a higher-order azimuthon.

Here, we would like to note that the Rabi oscillation is not feasible between arbitrary two azimuthons.
Only azimuthons with similar structures (e.g., the dipole and higher-order dipole azimuthons) can switch into each other,
and azimuthons with different symmetries (e.g., the dipole and quadrupole azimuthons) will not,
on the account that the overlap integrals in general areexactly zero, $\langle U_{m} V U_{n}\rangle = 0$.
We would like to note that the Rabi oscillation between two modes with opposite symmetry is also possible if the potential is anti-symmetrically modulated in the transverse plane \cite{zhang.prl.123.254103.2019}.

Generally, there is a frequency detuning $\ell=d-d'$ between the real modulation frequency $d'$ and the resonant frequency $d$.
Therefore, it is reasonable to have a look at the efficiency of the azimuthon conversion versus the detuning $\ell$.
However, one cannot obtain the direct efficiency via the projections of the field amplitude $\psi$ on the targeting azimuthons
due to the rotation of the azimuthons during propagation.
But the efficiency can be reflected by the Rabi oscillation period $z_R$ --- the bigger the value of $z_R$ the bigger the efficiency \cite{zhang.oe.25.32401.2017,zhang.lpr.12.1700348.2018,zhong.ol.44.3342.2019}.
The dependence of $z_R$ on frequency detuning $\ell$ is shown in Fig. \ref{fig3}(c).
As a result, one finds that the efficiency of the azimuthon conversion is the biggest at the resonant frequency, and
it reduces with the growth of the frequency detuning $\ell$.

\subsection{Square waveguide}

Now, we investigate the azimuthon transition in a square waveguide, which is generated by the potential in Eq. (\ref{fig2}) of the form
$
V(x,y) = V_0 \exp [-(x^{10}+y^{10})/w^{10}) ].
$
Again, we solve for the linear eigenmodes supported by the deep square waveguide, by using the plane-wave expansion method.
Connected with the geometry of the potential, the amplitude distributions of the linear modes are more
complex than those in the regular circular waveguide, therefore we denote them as the deformed modes.
In Fig. \ref{fig4}, we display three kinds of deformed modes and the corresponding azimuthons,
which will transform mutually, because of the relation $\langle U_{m} V U_{n}\rangle\neq0$.

\begin{figure}\centering
  \includegraphics[width=0.5\columnwidth]{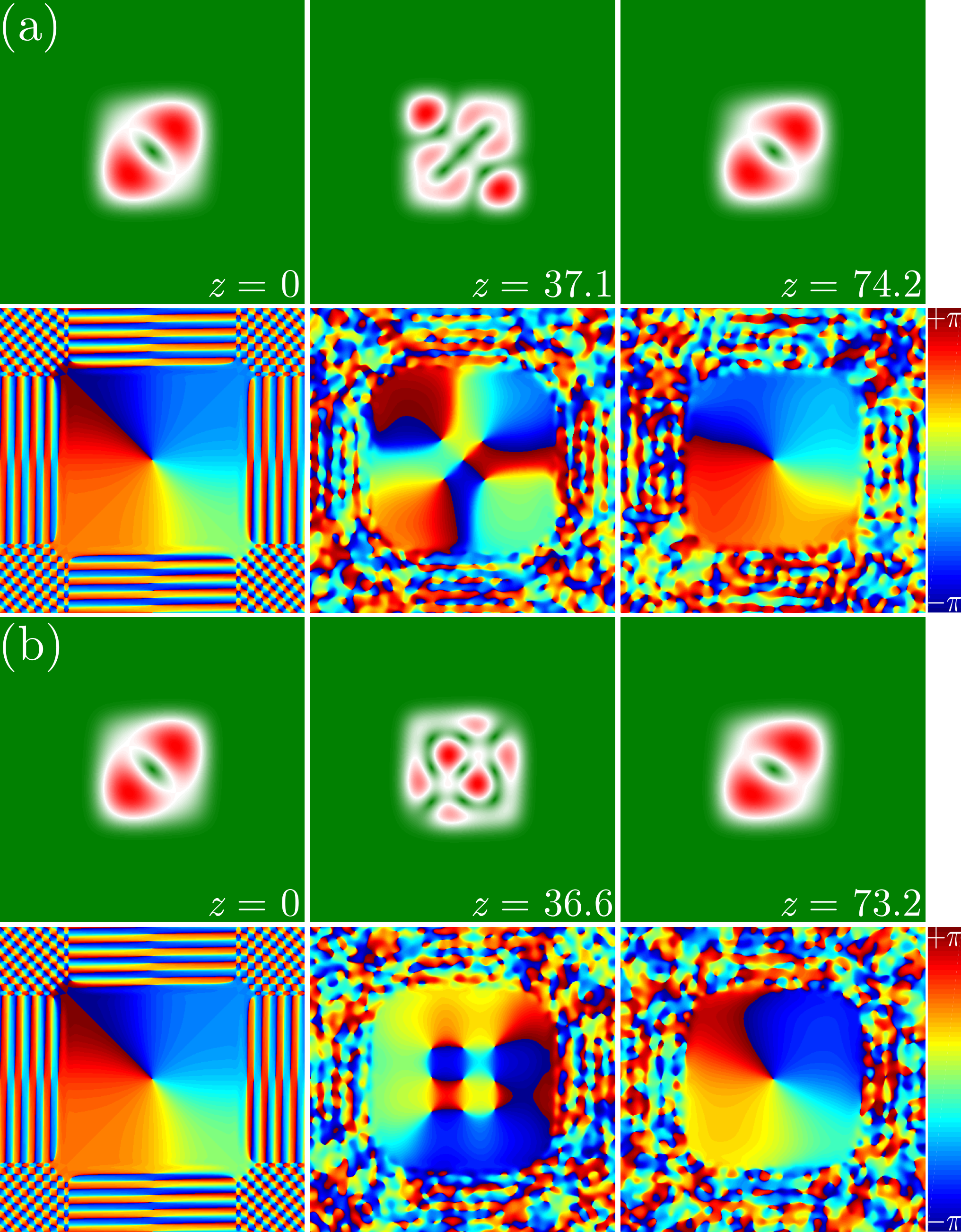}
  \caption{(a) Transition between dipole and hexapole azimuthons with $d\approx25.2$ and $\mu\approx0.085$.
  (b) Transition between dipole and hexapole azimuthons with $d\approx22.1$ and $\mu\approx0.034$.
  The weak longitudinally periodic modulation has to always exist during propagation.}
  \label{fig5}
\end{figure}

Different from the azimuthons in circular waveguides,
azimuthons in square waveguides rotate conditionally.
Due to the symmetry of the square waveguide, a rotating azimuthon will be deformed,
i.e., its profile changes.
Without considering nonlinearity, linear superposition of two degenerated modes [e.g., dipoles in Fig. \ref{fig4}(a),
and labelled as $u_{1,2}(x,y)$] is also dipole solution of the square potential, as  $u'_{1,2}(x,y) = [u_1(x,y) \pm u_2(x,y)]/\sqrt{\iint |u_1(x,y) \pm u_2(x,y)|^2 dx dy}$.
It has been found that if an azimuthon in a square waveguide can rotate it should meet the condition: $\mathcal{H}(\psi)>\mathcal{H}(u'_{1,2})$ with $\mathcal{H}$ representing the Hamiltonian \cite{zhang.oe.18.27846.2010}.
Even though the wobbling azimuthon can be established in our numerical simulations,
we are more interested in rotating azimuthons,
so we set again $A=0.4$ and $B=0.5$ in this section, to guarantee the rotation of azimuthons during propagation.
On the other hand, azimuthons will be deformed during propagation, because of the symmetry of the potential,
therefore in order to exhibit the azimuthon conversion in a more clear way,
one has to properly choose the value of $\mu$, to make the Rabi oscillation period almost equal to one half of the target azimuthon rotation period,
since the modulation frequency $d$ is determined beforehand.
Numerical simulations reveal that the rotation periods of the hexapole azimuthon and
the higher-order dipole azimuthon are $\sim 148.4$ and $\sim 146.4$, respectively.
Therefore, according to Eq. (\ref{eq13}), the modulation strength $\mu$ for the two cases should be $\sim0.085$ and $\sim0.034$, respectively.
The numerical demonstration of the Rabi oscillation is shown in Fig. \ref{fig5}.
Different from the setting for the case of circular waveguide, the longitudinal modulation always accompanies the square waveguide.

In Fig. \ref{fig5}(a), the hexapole azimuthon is obtained at $z\sim37.1$ (one half of the Rabi oscillation period,
and also a quarter of the rotation period of the hexapole azimuthon),
while in Fig. \ref{fig5}(b), the higher-order dipole azimuthon is obtained at $z\sim36.6$.
When the propagation distance reaches one Rabi oscillation period,
the dipole azimuthon is recovered with a small deformation.
To show the azimuthon conversion more transparently, we also display the corresponding phase distributions. Evidently,
there is only one phase singularity in the phase of the dipole azimuthon (the topological charge is 1),
five singularities for the hexapole azimuthon (the topological charge is 3),
and nine for the higher-order dipole azimuthon (the topological charge is again 1).
As seen, the phase distributions are in accordance with the expectations and with those displayed in Fig. \ref{fig4}.

\section{Conclusion}

We investigated and demonstrated Rabi oscillations of azimuthons in weakly nonlinear waveguides with weak longitudinally periodic modulations.
Based on the coupled mode theory, we find the period of the Rabi oscillation,
which is affected by the modulation strength and also by the spatial symmetry of the azimuthon.
The analysis is feasible for both circular and square waveguides,
and can be extended to waveguides with other symmetries.

Based on the model taken in this work, switching between a vortex-carrying azimuthon and a multipole that is free-of-vortex will not happen.
The reason is that the initial azimuthon is composed of two degenerated modes $u_{1,2 }$,
which will switch into another two degenerated modes $u_{3,4}$ during propagation.
So, the output is a composition of $u_{3,4}$ which carries vortex.
However, if the potential is modulated transversely in a proper manner,
such a switch becomes possible since both the longitudinal and transverse phase matching can be satisfied.

\section*{Acknowledgements}
This work was supported by
Guangdong Basic and Applied Basic Research Foundation (2018A0303130057),
National Natural Science Foundation of China (U1537210, 11534008, 11804267), and
Fundamental Research Funds for the Central Universities (xzy012019038,
xzy022019076). M.R.B acknowledges support from the NPRP 11S-1126-170033 project from the Qatar national Research Fund, while K.C.J. and Y.Q.Z. acknowledge the computational resources
provided by the HPC platform of Xi'an Jiaotong University.

%

\end{document}